\documentclass[twocolumn]{aastex63}
\usepackage[utf8]{inputenc}
\submitjournal{ApJ}

\newcommand{\msun}{\ifmmode {\,{\rm M}_{\odot}}\else${\,{\rm M}_{\odot}}$\fi}
\newcommand{\Rsun}{\ifmmode {R_{\odot}}\else${M_{\odot}}$\fi}

\begin{document}

\title{Mass Matters: No Evidence for Ubiquitous Lithium Production in Low-Mass Clump Giants}

\correspondingauthor{Julio Chanam\'e}
\email{jchaname@astro.puc.cl}

\author[0000-0003-2481-4546]{Julio Chanam{\'e}}
\affiliation{Instituto de Astrof\'isica, Pontificia Universidad Cat{\'o}lica de Chile, Av. Vicu{\~n}a Mackenna 4860, 782-0436 Macul, Santiago, Chile}

\author[0000-0002-7549-7766]{Marc H. Pinsonneault}
\affiliation{Department of Astronomy, The Ohio State University, 140 W. 18th St., Columbus, OH 43210, USA}

\author[0000-0002-9052-382X]{Claudia Aguilera-G\'omez}
\affiliation{N\'ucleo de Astronom\'ia, Universidad Diego Portales, Ej\'ercito 441, Santiago, Chile}

\author[0000-0002-7549-7766]{Joel C. Zinn}
\altaffiliation{NSF Astronomy and Astrophysics Postdoctoral Fellow.}
\affiliation{Department of Astrophysics, American Museum of Natural History, Central Park West at 79th Street, New York, NY 10024, USA}

\begin{abstract}

Known sources of lithium (Li) in the universe include the big bang, novae,
asymptotic giant branch stars, and cosmic ray spallation. During their
longer-lived evolutionary phases, stars are not expected to add to the Li
budget of the Galaxy, but to largely deplete it.  In this context, recent
analyses of Li data from GALAH and LAMOST for field red clump (RC) stars
have concluded that there is the need for a new production channel of Li,
ubiquitous among low-mass stars, and that would be triggered on the upper
red giant branch (RGB) or at helium ignition. This is distinct from the
“Li-rich giant” problem and reflects bulk RC star properties. We provide
an analysis of the GALAH Li data that accounts for the distribution of
progenitor masses of field RC stars observed today. Such
progenitors are different than today’s field RGB stars. Using standard
post main-sequence stellar evolution, we show that the distribution of Li
among field RC giants as observed by GALAH is consistent with standard model
predictions, and does not require new Li production mechanisms. Our model
predicts a large fraction of very low Li abundances from low mass
progenitors, with higher abundances from higher mass ones.  Moreover,
there should be a large number of upper limits for RC giants, and higher
abundances should correspond to higher masses. The most recent GALAH data
indeed confirm the presence of large numbers of upper limits, and a much
lower mean Li abundance in RC stars, in concordance with our interpretation.

\end{abstract}


\section{Introduction} \label{sect:intro}

Lithium (Li) is a particularly important chemical element for astrophysics, with an impact on problems ranging from the early universe to the assembly of planetary systems recently formed around young stars.  The $^7$Li isotope is one of the few species created during big bang nucleosynthesis (BBN), and it burns in stellar interiors via proton capture at relatively low temperatures.  Because of its fragility, stars with outer convective envelopes easily deplete any Li on their surfaces, making it a useful thermometer of the physical conditions that set the controls for the heart of the Sun and the stars. 

Besides BBN, other known astrophysical sources of Li synthesis include asymptotic giant branch stars \citep{cameron71,sackmann92}, novae \citep{novae1979,tajitsu2015,izzo2015} and cosmic ray spallation \citep{spallation1970,spallation1992}.  During their longer-lived evolutionary phases, stars in general are not expected to add to the Li budget of the Galaxy, but to largely deplete it.

Large spectroscopic surveys are now providing detailed abundance patterns for hundreds of thousands of stars. GALAH \citep{galah} and LAMOST \citep{LAMOST}, among others, include Li data, and some recent works include \cite{smiljanic2018,casey2019,gao2019,gao2021}; and \cite{martell2021}.  In this paper we focus on GALAH Li data for evolved stars.

Interpreting Li data is not straightforward. There is a large literature on Li abundances in stars; for example, see \citet{sestito} for a discussion in the open cluster context, and \cite{Lidesert} for a recent analysis of field stars. The known observational pattern is complex, but it is useful to summarize the main expectations for evolved stars (see \citealt{pinsono1997} for a more detailed discussion).

Solar-type and lower mass stars can destroy a significant amount of Li during the pre-main sequence \citep{iben1965}; this depletion may be modified by the structural effects of star spots and magnetic fields \citep{somers2016}.  Stars also experience main sequence (MS) depletion (e.g., \citealt{sestito}); this depletion depends on mass, composition, and age, and there is a dispersion in depletion even between open cluster stars with the same mass. The latter provides strong evidence for rotationally induced mixing as an explanation for Li depletion, as stars that are born with different rotation rates mix at different rates \citep{pinsono1989}.

There is a Li dip, located at the transition from radiative to convective envelopes on the MS \citep{Lidip}; stars in this domain almost completely destroy Li on the MS, and this phenomenon is clearly seen in GALAH data \citep{gao2020_Lidip}. Li is difficult to measure in upper MS stars, but there is some evidence for dispersion and MS destruction from observations of red clump (RC) stars in young open clusters \citep{gilroy1989}. Once stars leave the MS, they develop deep convective envelopes and dilute their Li content.  Li on the red giant branch (RGB) can further be reduced by extra mixing on the upper RGB (e.g., \citealt{shetrone2019}); mass transfer can induce severe Li depletion \citep{ryan2001}; and under some rare circumstances, Li can be produced in evolved stars \citep{cameron71,sackmann92}.

In this context, a recent work by \cite{nature} claimed the discovery of evidence for a new source of Li production that would be ubiquitous to low-mass stars and which would be associated to some event occurring somewhere in between the tip of the RGB and the subsequent helium (He) burning phase.  Possibilities for the underlying mechanism behind the new production channel have since been proposed (e.g., \citealt{flash,neutrinos}).  Along similar lines, an independent analysis of field giants with Li data from LAMOST and stellar masses from {\it Kepler} and {\it K2} also argued for evidence of Li production during the helium flash \citep{zhang2021}.

In GALAH data, the Li abundances observed among core He-burning, or RC stars are higher than those observed in shell hydrogen(H)-burning, upper RGB stars. However, as noted above, Li is subject to strong mass and composition dependent destruction mechanisms. Selection effects are therefore a serious concern, and, in order to use the relative RC and RGB Li abundances as evidence for production, it would be necessary to establish that the distribution of Li detections is representative of the underlying population, and that the RC stars are the evolutionary successors of the RGB stars.


In this work we demonstrate that such population effects are important for interpreting Li data, and that non-detections are a serious concern. We conclude that the case for universal Li production has not been established, and that further analysis, especially including the large cohort with upper limits, is required.  The GALAH RC giants are part of a field population, and, unlike the giants in a stellar cluster, span a wide range of stellar mass and composition.  This is crucial for the subject at hand because low mass stars experience much more severe MS depletion than high mass stars.  The same happens on the RGB, where low-mass stars deplete Li much more efficiently than stars just slightly more massive.  When comparing field populations, it is therefore of paramount importance to ensure that differences in the underlying mass distributions are accounted for.  This complication is avoided if looking at stellar clusters, and indeed a recent analysis of cluster data from the Gaia-ESO survey cannot confirm any universal Li production event between the upper RGB and the RC \citep{magrini2021}.

Specifically, in a field population such as that of GALAH, red giants and clump giants do not come from the same MS progenitors.  Clump giants are a younger population than red giants, because the lifetime of the RC is much less mass dependent than the lifetime of the RGB.  And since stars of different mass experience different amounts of Li depletion during the MS (as well as during post-MS), then also the initial conditions of Li content at the end of the MS were not the same for the RGB and RC field populations observed today.

A second conclusion reported in \cite{nature}, but which will not be further addressed in the present work, is that the widely used threshold of Li abundance, A(Li) = 12. + $\log(N_{\rm Li}/N_{\rm H}) = 1.5$ dex, above which giants have been historically classified as Li-rich should be revised downward.  In reality, as results of the present work will also serve to show, that threshold is dependent on stellar mass, as was first reported by \cite{claudia2016,trumpler2016}, based on a study of Li enrichment in RGB stars arising from the engulfment of substellar companions. 


\begin{figure*}
    \centering
    \includegraphics[width=17.5cm]{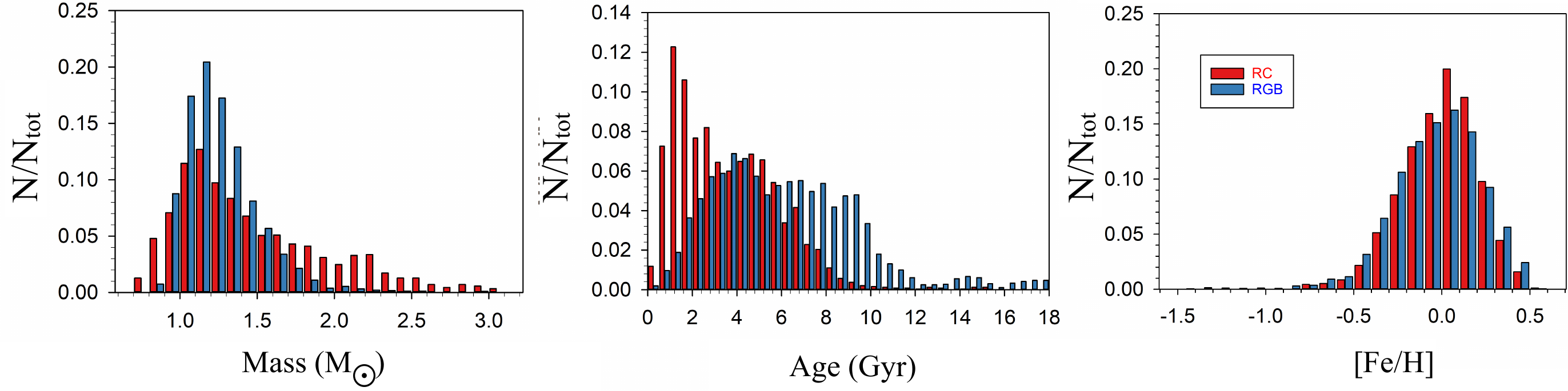}
    \caption{The distributions of stellar mass, age, and metallicity for field RGB and RC stars (blue and red, respectively) in the {\it Kepler} field as determined by APOKASC.  As a population, field RC stars are somewhat younger and more massive than field RGB stars.  See text for details and discussion. 
    }
    \label{fig:APOKASC}
\end{figure*}

This paper is organized as follows.  In Section \ref{sect:samples} we discuss the properties of the GALAH populations of red giants and clump giants, where we assume that they can be well represented by the corresponding populations in the {\it Kepler} field as characterized by the Second APOKASC Catalog \citep{apokasc}.  This allows us to take advantage of the availability of seismic mass and age determinations, for RGB and RC stars.

In Section \ref{sect:model} we describe and discuss the results of our model for simulating the Li content of the {\it Kepler}/GALAH clump giants, starting from adequate initial conditions, and in Section \ref{sect:conclusions} we summarize our results and conclusions.

\section{Clump giants in GALAH and {\it Kepler}} \label{sect:samples}

In order to best quantify the differences between the field populations of RGB and RC stars in general, it would be convenient to rely on stars with seismic characterization.  Unfortunately, the {\it Kepler} field is not in the same hemisphere as the GALAH footprint, and the overlap between GALAH and {\it K2} is small.  Our analysis (Section \ref{sect:model}) will therefore combine seismic masses for stars in the {\it Kepler} field with Li abundances obtained for stars towards another (large) region of the Galaxy, effectively ignoring known differences between the stellar mass distributions of the two samples, which most likely arise from Galactic structure (e.g., \citealt{miglio2013,sharma2016}).  In particular, populations in the \textit{Kepler} field are dominated by the thin disk; but the field was chosen to avoid the disk mid-plane, and stars below 1 Gyr are under-represented.  We briefly discuss some potential consequences in the conclusions (\S\,\ref{sect:conclusions}).

\begin{figure}
\centerline{\includegraphics[width=1.0\columnwidth]{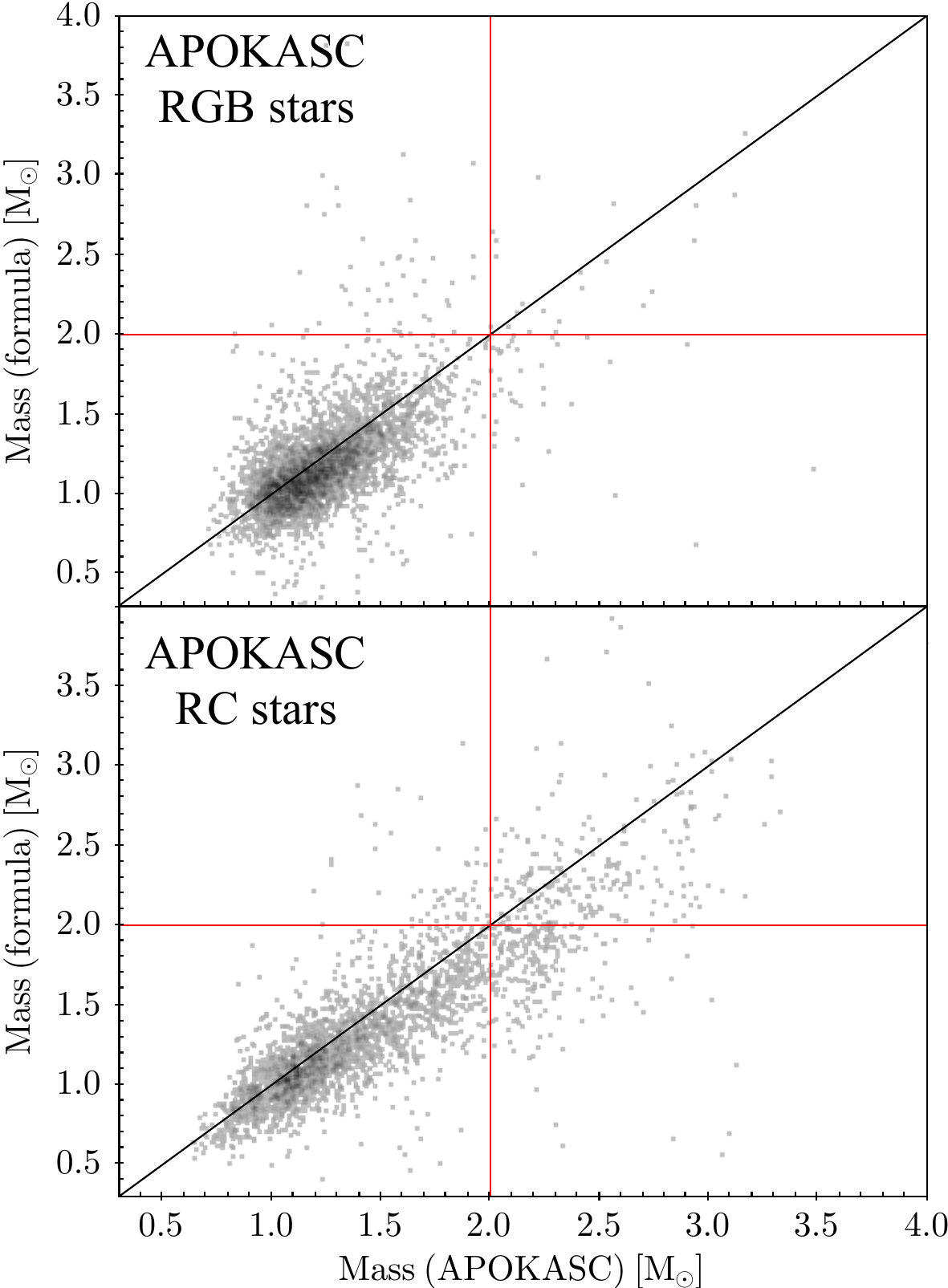}}
    \caption{Comparison between stellar masses determined by stellar parameters (Equation \ref{eq:mass}) and masses determined from asteroseismic measurements, for the APOKASC samples of red and clump giants.  It can be seen that, for the RC sample, selection of stars less massive than 2.0\msun\, following the mass determination via stellar parameters results in including a number of RC giants with seismic masses up to 2.5\msun\, and beyond.}
    \label{fig:formula}
\end{figure}

We take advantage of the Second APOKASC Catalog \citep{apokasc}, which provides reliably determined evolutionary states, surface gravities, mean densities, masses, radii, and ages for more than 6,000 evolved stars.  Figure \ref{fig:APOKASC} shows the mass, age and metallicity distributions of RGB and RC stars in APOKASC2.  From the full sample in \citet{apokasc} we removed stars with low quality seismic solutions or ambiguous evolutionary states.  High age estimates (i.e., older than the age of the universe), and their corresponding low stellar mass estimates (i.e., lower mass than that capable of evolving off the MS in a Hubble time, which we approximate as $\sim 0.8\msun$), either reflect random errors from lower true ages/higher true masses, or arise from stars that experienced binary interactions.  We have retained these so as to avoid biasing the mean properties that we are interested in identifying, so they appear in Figures \ref{fig:APOKASC} and \ref{fig:formula}, but have been removed from the final results in Figure \ref{fig:model}.

The two mass distributions peak above 1\msun, at 1.21 and 1.32\msun, respectively.  Similar masses were reported by \cite{nature} for the GALAH clump giants.  However, the RC distribution appears significantly skewed towards higher masses, while the RGB distribution much less so.  This suggests that the RC population must be composed, in general, of stars younger than an old RGB star of 1\msun\, (the model chosen by \citealt{nature} for their comparison to the GALAH clump giants).  Examining the middle panel of Figure \ref{fig:APOKASC}, one can see that indeed this is the case, with the age distribution of RC stars in APOKASC being $\sim 2-3$ Gyr younger than that of RGB stars.

Given the severe known mass sensitivity of Li depletion, even 0.1\msun\, offsets can be a major concern. In addition, we expect very different Li depletion/destruction properties for high and low mass stars. The fraction above 1.5 solar masses is 14\% for the RGB and 38\% for the RC. Given the strongly peaked RGB mass distribution, even these estimates are likely to be conservative: the lifetime on the RGB for massive stars is much less than that for low mass stars, and, as the latter are more numerous, the underlying population is skewed.  Therefore, many of the high mass estimates for RGB stars are likely to arise from random errors in ordinary low mass giants.

A potential additional concern is that today's RGB population and the RGB progenitors of today's RC population have different metallicity distributions as well, and extra mixing on the RGB is known to be metallicity-dependent \citep{shetrone2019}.  However, the rightmost panel of Figure \ref{fig:APOKASC} shows that the distributions of the two populations are basically indistinguishable above [Fe/H] $ > -1$ (although we don't show it, the lowest-mass RGB stars are found at the lower metallicities, some of which are likely associated with the halo).  Therefore, while metallicity-dependent mixing may impact Li depletion, it would operate similarly in both RGB and RC populations (i.e., in the RGB progenitors of the latter), and should not be expected to contribute to selection effects in observed Li abundances.


A key aspect in the \cite{nature} analysis is the selection of the sample of RC giants.  These are selected directly from the Hertzprung-Russell (HR) diagram, restricting the analysis to stars less massive than 2\msun.  Stellar masses are estimated from the relation
\begin{equation}
    \log(M/\msun) = \log(L/L_{\odot}) + 4\log(T_{\rm eff}^{\odot}/T_{\rm eff}) + \log(g/g_{\odot}),
\label{eq:mass}
\end{equation}

\noindent where the luminosities are taken from Gaia DR2 \citep{dr2}, effective temperatures ($T_{\rm eff}$) and surface gravities are from GALAH DR2, and $T_{\rm eff}^{\odot} = 5,777$ K and $\log g_{\odot} = 4.44$ are the adopted solar parameters.  Finally, using asteroseismic parameters taken from the literature for a fraction of their RC sample, \cite{nature} estimate a contamination from first ascent RGB stars of $\sim 10$\%.

In Figure \ref{fig:formula} we compare the stellar masses as computed by \cite{nature} from Equation \ref{eq:mass}, against asteroseismic determinations, always for the RGB and RC populations in the {\it Kepler} field.  While there is a scatter of $\sim 0.3-0.4$\msun\, throughout, in both cases there is a relatively decent one-to-one correspondence between both mass determinations, with possibly a small systematic in the sense of slightly larger values for the seismic masses.  However, more relevant to our discussion is the fact that, as the lower panel of Figure \ref{fig:formula} shows, a selection of RC giants less massive than 2\msun\, based on a stellar mass computed from the formula in Equation \ref{eq:mass} leads to the inclusion of a non-negligible fraction of stars that are actually more massive than 2\msun.  That is, when \cite{nature} attempt to cut their RC sample at 2\msun, what really happens is that stars as massive as 2.5-3\msun\, creep into the sample.  As we show in the next section, for clump giants as massive as these, it is perfectly natural to expect Li abundances as observed in the GALAH data, which cannot be compared to what is expected for an old red giant of 1\msun.

\section{Standard model prediction for the Li content of field clump giants} \label{sect:model}

In order to predict the Li abundances of a given population of RC stars we need three ingredients: (1) the stellar mass distribution of the progenitors of the RC stars, (2) initial conditions for the Li content on the MS as a function of stellar mass, and (3) a stellar evolution model.  The latter is then used to forward model the Li initial conditions up to the RGB tip, following the stellar mass distribution assumed for the progenitors.  Given that the main goal of this work is to determine whether new physics is needed regarding the behavior of Li in low-mass giants (i.e., whether there is any need or not for new, unknown Li production/destruction mechanisms), we assume that the Li abundance computed at the RGB tip is the same as that on the horizontal branch.  Comparison to the observational data will then decide whether this assumption needs to be revised or not.

The only stellar interiors ingredient that determines the outcome of this exercise is the ability of a star of a given mass to deplete Li during their post-MS evolution, so we discuss our stellar models first.

For a given MS progenitor mass and an initial Li abundance, we use standard stellar models constructed with the Yale Rotating Evolutionary Code (YREC; \citealt{pinsono1989,demarque2008}) to compute the depletion of Li from the MS turnoff until the tip of the RGB.  We stress that we use standard stellar models, i.e., with no extra- nor thermohaline mixing, and where the only mixing agent is convection.  

We will assume, for simplicity, solar metallicity for all the stars in our models.  This assumption is justified by two reasons.  First, the distribution of metallicities of field RC stars in APOKASC peaks at solar metallicity, as shown in the right-most panel of Figure \ref{fig:APOKASC}.  Possibly more relevant, the stars analized by \citet{nature} also have a metallicity distribution that peaks almost at solar metallicity ($-0.1$ dex). Second, the main point of our exercise is to quantify the impact of mass on the predicted stellar Li depletion.  A full simulation would account for chemical evolution, and both the mass and metallicity dependence of Li depletion. This would require a complete simulation of the relevant field populations, a drastic increase in scope.  However, as we will see below, the mass effects alone are dramatic enough to drastically impact the interpretation of Li data.


\begin{figure}
\centerline{\includegraphics[width=1.0\columnwidth]{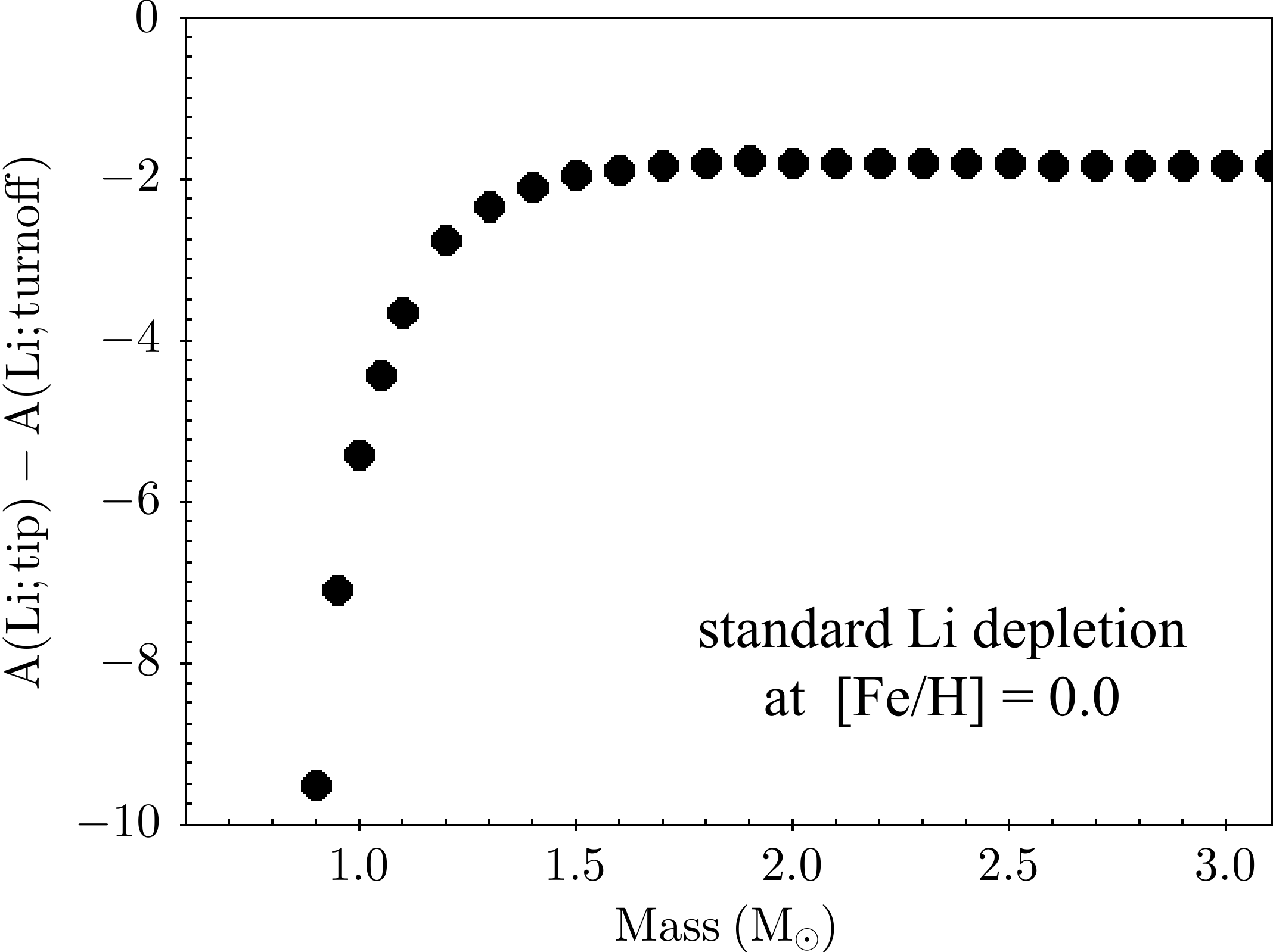}}
    \caption{Post main sequence depletion of lithium as a function of stellar mass, for standard evolutionary models (i.e., mixing only due to convection) with solar metallicity.  Note the largely different depletion factors between stars of 1 and 2\msun, of about 7 and 2 dex, respectively.}
    \label{fig:depletion}
\end{figure}

In Figure \ref{fig:depletion} we show the amount of Li depleted by our models as a function of stellar mass.  This will be a good approximation for both standard models and ones with rotationally induced mixing. Two effects in rotating models that could affect the picture of Li evolution shown in Figure 3 are: 1) increasing the surface Li by mixing up previously gravitationally-settled Li; and 2) destroying additional Li as more Li is mixed to hot enough temperatures for Li destruction.  Regarding the first effect, data from star clusters indicate that, at least for the Li dip, there is no such dredge-up of a sub-surface reservoir \citep{balachandran1995}.  Lithium on the MS therefore is largely destroyed, rather than simply being hidden below the surface by gravitational settling.  To the extent that the true Li content for field MS stars is higher than their surface abundances would indicate, then we will be under-estimating the true Li abundances in evolved stars, so our approach is conservative.

For our initial conditions on Li abundance on the MS, we go to the same source as \cite{nature} and use data from GALAH DR2 to set the distribution of abundances for stars leaving the MS.  Note that Li measurements in GALAH DR2 are not flagged as either detections or upper limits, even though it is to be expected that a fraction of those measurements are formally upper limits.  The Third Data Release (DR3) of GALAH \citep{galahDR3} flags upper limits, but we chose to stay with the unflagged DR2 measurements because using the more complete characterization of the Li data from DR3 would amount to an important difference with respect to the analysis of \cite{nature}, hence potentially diluting one of the main points of the present work, which is about the properties of the field population.

\begin{figure}
\centerline{\includegraphics[width=1.0\columnwidth]{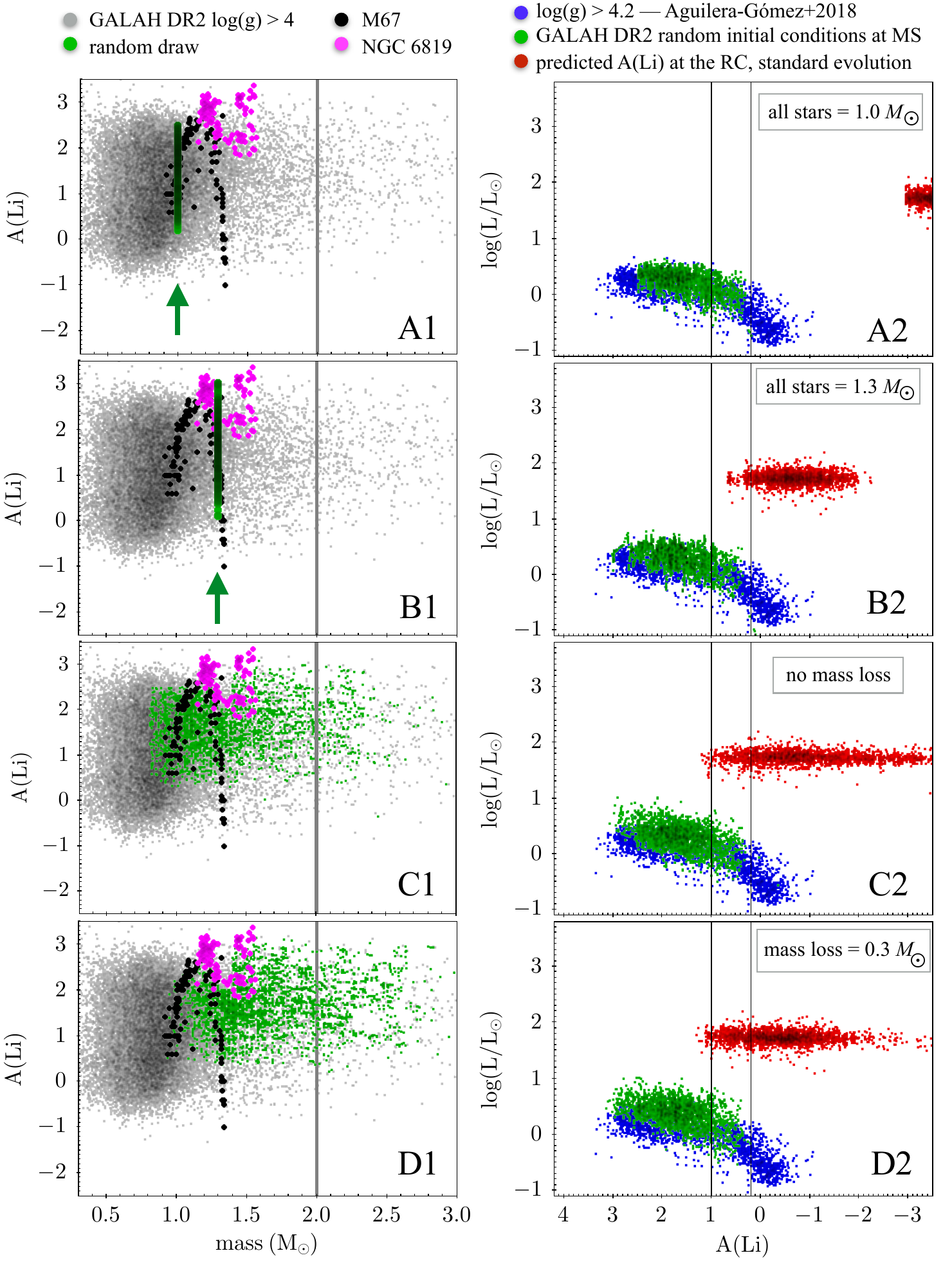}}
    \caption{Initial conditions (ABCD-1) and results (ABCD-2) of our simulations, for the four experiments discussed in this paper.  Cases A and B illustrate the situation when a single stellar mass is adopted for the progenitors of field RC stars (green arrows aid in signaling location of progenitors for these two cases), while cases C and D adopt more realistic distributions of initial conditions for the RC progenitors, without and with mass loss, respectively.  {\bf LEFT}: Abundances of Li on the MS as a function of stellar mass, used as initial conditions for our simulations.  In gray, GALAH Li measurements for stars with $\log$ g\,\,$> 4$ and errors smaller than 0.04 dex. Data for stars in M67 ($\sim 4$ Gyr) and NGC 6819 ($\sim 2$ Gyr) are shown in black and magenta, respectively, and are used in order to avoid initial Li abundances that would be appropriate for very young stars or the region of the Li dip (see text). In green, the randomly selected initial conditions for each simulation, which correspond to real stars in GALAH DR2, as described in the text.  {\bf RIGHT}: In red, the predicted Li abundances for field RC stars for the corresponding simulation.  The vertical lines indicated the location of the bulk of the RC stars observed by GALAH.  Green points are the same as in the left panels, and blue point are the Li data from \cite{Lidesert} for MS stars, used here just as background upon which to see how the (green) initial conditions change between our experiments.}
    \label{fig:model}
\end{figure}

We need to restrict ourselves to the mass range of MS stars that are young enough to reach the RC and RGB, and old enough to be close to the end of their MS lifetime. We therefore restrict our sample to stars with $\log$ g $> 4$, and errors in A(Li) smaller than 0.04 dex, which we deem as a secure limit for good measurements based on the inspection of the distribution of the overall GALAH DR2 Li data.  The run of Li initial conditions on the MS as a function of stellar mass is shown as gray dots in all the left hand panels of Figure \ref{fig:model} (ABCD-1).  These are the actual GALAH DR2 data for stars with $\log$ g $> 4$ and small errors, as mentioned above.  At any given mass, there is a distribution of possible Li abundances on the MS, which we will account for in our model predictions by directly taking real Li measurements from these mass-dependent distributions.  Another useful sample for obtaining initial conditions for Li on the MS is that of \cite{Lidesert}, shown in blue in the right-hand panels of Figure \ref{fig:model} (ABCD-2), but we use this sample only as a background to show the run of initial conditions of our experiments below.  Our luminosities are taken directly from Gaia DR2, which reports estimates of stellar parameters, extinction, reddening, and luminosities from the Apsis data processing pipeline \citep{apsis}.

Note that stars from GALAH, which we use to obtain initial conditions for our simulations, are not all located in the MS turnoff, and include stars in earlier MS stages than the turnoff. The use of this sample directly would not account for possible MS depletion and would overestimate the Li in the simulated stars.  We account for this by using additional cluster data to estimate the maximum Li expected for stars of different masses after they experience depletion.  We consider as a reference data for M67 \citep{m67} and NGC 6819 \citep{ngc6819}, with $\sim 4$ and $\sim 2$ Gyr, respectively, included in Figure 4.  We treat these as Li upper envelopes to our turnoff distribution, i.e., we reject any draw in our experiments above these values, lowering the Li abundance for masses in and below the Li-dip, but preserving Li of stars of higher mass that do not experience significant Li depletion.  Notice also that the descendants of these higher mass stars are more common in the clump (Figure 1), which would explain why they have high Li relative to what we expect from lower-mass stars.
Then, our recipe for initial Li distribution in the RGB uses the GALAH dataset with additional conditions from these clusters to account for MS depletion: if $\mathrm{M/M_\odot}<1.3$: reject draws with A(Li)\,$>2.5$; if $1.3<\mathrm{M/M_\odot}<1.4$: reject draws with A(Li)\,$>2.0$; and if $\mathrm{M/M_\odot}>1.4$: allow all draws.  In practice, our experiments show that there are enough progenitors of field RC stars with M $> 2$\msun\, that the above recipe has little effect on the final result.

We perform four experiments, using four different sets of assumptions.  Our whole exercise is to simply take a given run of initial conditions of Li content as a function of stellar mass, and predict the distribution of Li among RC stars using evolutionary models.  The four cases are illustrated on four corresponding pairs of panels in Figure \ref{fig:model}, labeled A through D.  The left-hand panels (A1 through D1) show (in green) the selection of initial conditions of each experiment, i.e., the distribution of progenitor stellar masses and their starting Li abundances at the MS turnoff.  The right-hand panels (A2 through D2) show the result of each experiment, i.e., the initial (green, blue) and final (red) distributions of Li abundances, at the luminosity of the MS turnoff and the RC, respectively.  The two vertical lines indicate the range of Li abundances of the bulk of the RC stars as measured by GALAH.  That is, one can consider that predictions are successful insofar as enough red points fall between these lines.

First, if we assume that the RC population can be well represented by progenitor stars of 1\msun\,(A1), the resulting Li abundances would be all very low, very far from the observed levels, as shown in panel A2 in Figure \ref{fig:model}.  This is what led \cite{nature} to announce the discovery of Li production common to all low-mass stars.  In order to illustrate how critical the adopted mass of the progenitors is for this problem, in panels B1-B2 of Figure \ref{fig:model} we run our simulation assuming that the RC progenitors are all of 1.3\msun\, instead, i.e., just 0.3\msun\, more massive than adopted by \cite{nature}.  As can be seen, the impact is very large, with the bulk of red points ending at higher final Li abundances than the previous case, and a small fraction of predicted RC stars now reaching the observed levels of Li in the GALAH sample.  Once again, this is naturally expected, because an RGB star of 1.3\msun\, depletes Li much less efficiently than one of 1.0\msun, as illustrated in Figure \ref{fig:depletion}.  Conversely, although we don't show it as an example, adopting even slightly higher masses would put progenitor stars right in the Li dip (the run of magenta points from NGC 6819), and our simulation would predict only upper limits at the RC. The pattern would reverse above the Li dip, with higher mass turnoff stars having higher initial Li.

Next, we perform the exercise accounting for more realistic initial conditions.  We assume the stellar mass distribution of field clump giants observed by GALAH to be well represented by that of the {\it Kepler} sample (see \S\,\ref{sect:samples}).  Then, for each RC star in APOKASC, we follow the same procedure as \cite{nature} and compute the stellar mass from the stellar parameters according to Equation \ref{eq:mass}.  If the resulting mass is larger than 2\msun, we do not include the star and move to the next one in the list.  If the resulting mass is smaller than 2\msun, we keep the star but from now on we rely on its seismic mass determination from APOKASC.  As shown in the bottom panel of Figure \ref{fig:formula}, this procedure is permitting RC stars more massive than 2\msun\, to make it into the final sample.

Note that, because of mass loss during the upper RGB, the mass of the RC stars as selected above must be smaller than the mass of their MS progenitors.  In order to account for the effect of mass loss on our model predictions, we run our procedure for two cases, meant to illustrate limiting possibilities.  First, a case with no mass loss at all, so that the mass of the RC star as selected above is the same as the mass of its progenitor at the MS turnoff.  And second, a somewhat extreme case in which all stars lose 0.3\msun\, in the upper RGB, so that the mass of the MS progenitor at the turnoff is 0.3\msun\, larger than that of the RC star as determined earlier.  Mass loss shifts the MS mass distribution of the RC stars to even higher mass, as the current mass is an underestimate of the mass at earlier stages.  Panels C1-C2 in Figure \ref{fig:model} illustrate our results for the case with no mass loss, while panels D1-D2 correspond to the case where all stars lose the same amount of mass as just described.

Next, for each RC star in APOKASC (with a mass smaller than 2\msun\, according to Equation \ref{eq:mass}), we assign a randomly drawn initial Li abundance that follows the observed distribution of A(Li) as described above for MS stars at the corresponding MS progenitor mass.  See panels C1 and D1 of Figure \ref{fig:model}, for the initial conditions for the cases without and with mass loss, respectively.  It can be seen that the mass distribution of the MS progenitors for the case including mass loss is shifted to higher values exactly by 0.3\msun, as designed.

\begin{figure}
\centerline{\includegraphics[width=1.0\columnwidth]{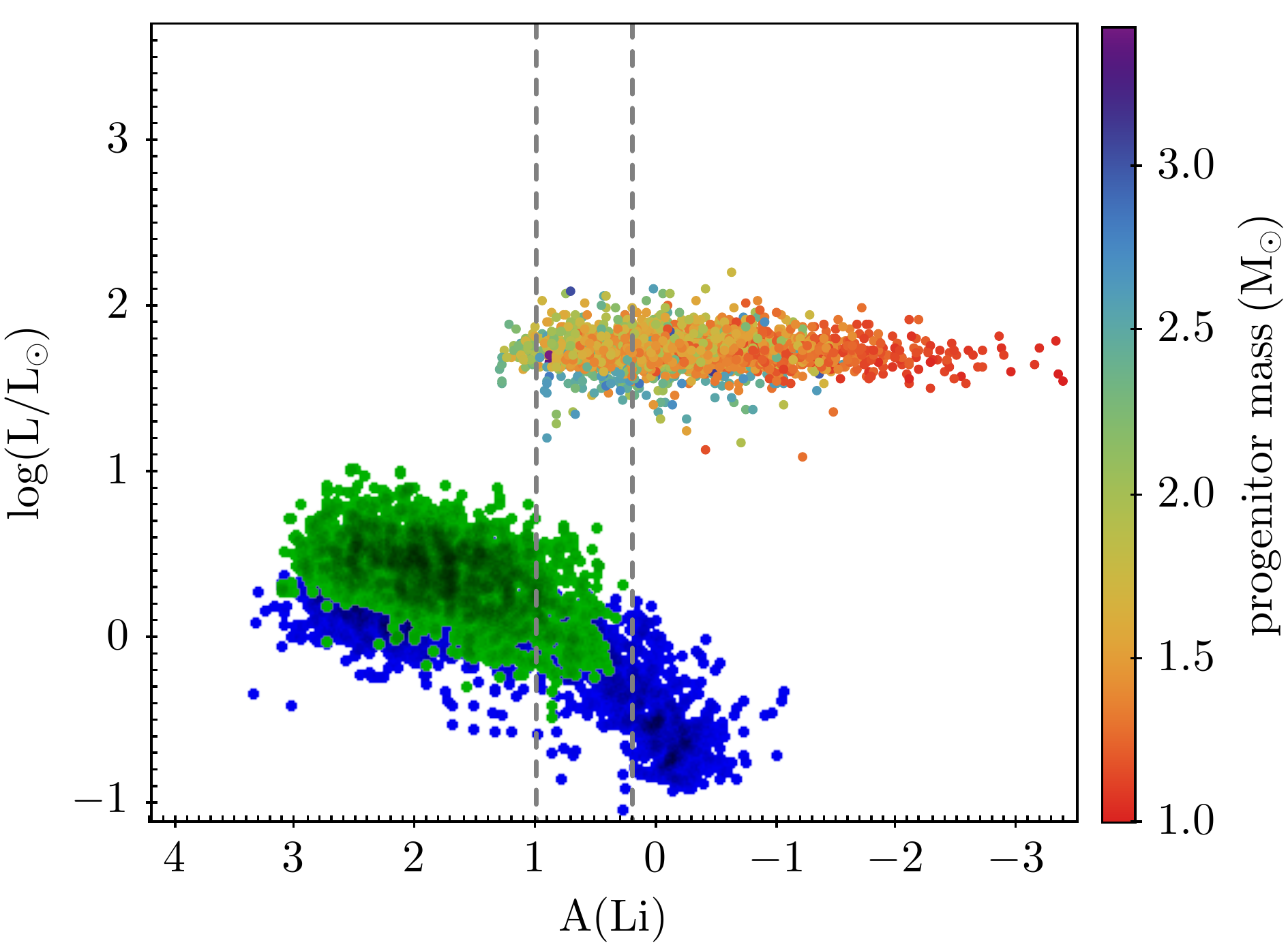}}
    \caption{Same as panel D2 of Figure \ref{fig:model}, with the simulated RC color coded according to the mass of the progenitor stars.  One solar mass progenitors populate the very low Li region of the RC, with A(Li) $< -1$, while the region of the RC populated by GALAH DR2 detections (region between the vertical dashed lines) is explained by higher mass progenitors, with M$\,\,\sim 1.5 \msun$ and above.}
    \label{fig:mass_code}
\end{figure}

The results of forward modeling these initial conditions are depicted in panels C2 and D2 of Figure \ref{fig:model}, for the case without and with mass loss, respectively.  These should be compared with the similar diagram in \cite{nature} (their Fig.3).  It can be seen that our standard model predictions, accounting for the range of stellar masses of the progenitors of clump giants, cover the region occupied by the data quite well, regardless of the assumption on mass loss.   Moreover, in Figure \ref{fig:mass_code} we show an enlarged version of panel D2 of Figure \ref{fig:model} that color codes the simulated RC according to progenitor's mass, therefore showing the stellar mass distribution of the field RC as predicted by our simulation.  Figure \ref{fig:mass_code} shows clearly that $1\msun$ progenitors produce RC giants with low levels of Li, A(Li) $< -1$, while the RC region as observed by GALAH, with A(Li) $\sim$ 0.5 - 1.0 dex, is expected to be populated by somewhat higher mass progenitors, with M $\sim 1.5\msun$ and above, such as we see in the \textit{Kepler} RC (Figure \ref{fig:APOKASC}, left panel).  This prediction is testable with asteroseismic data.

\begin{figure}
\centerline{\includegraphics[width=1.0\columnwidth]{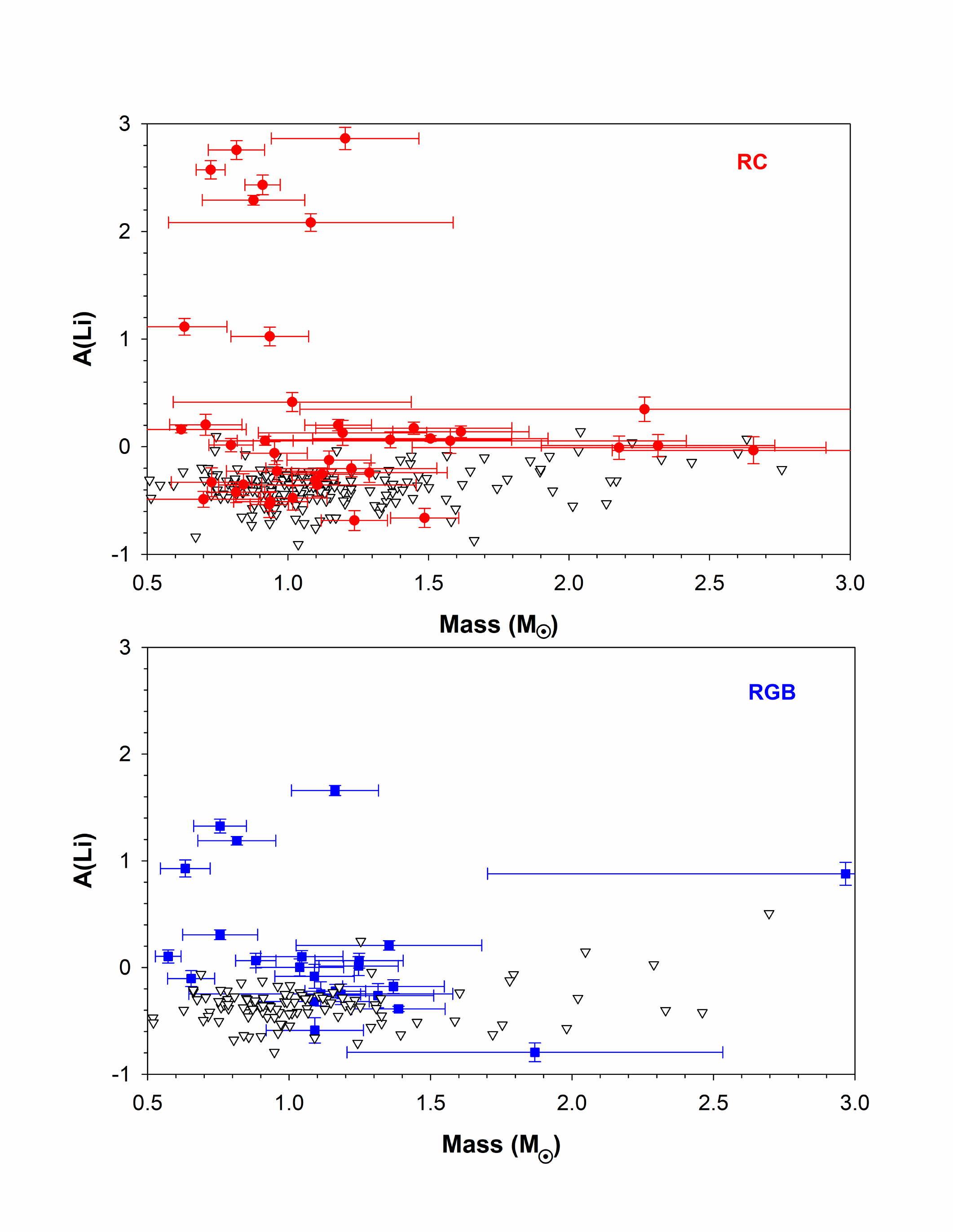}}
    \caption{Lithium measurements with {\it K2} asteroseismic data for stars classified as red clump (top panel) and red giant (bottom panel). Red clump and red giant branch detections are the blue (top) and red (bottom) symbols, while upper limits are denoted by triangles. Masses are taken from the \citet{zinn2021} asteroseismic factors with \citet{galahDR3} GALAH temperatures; Li measurements and limits are taken from \citet{galahDR3}. Evolutionary states are from \citet{zinn2021}.}
    \label{fig:K2_galahdr3}
\end{figure}

The simulations in C2 and D2, besides naturally explaining the observed bulk distribution of Li in the RC, also predict an extended tail of RC stars with very low Li abundances, which is not seen in the GALAH data analyzed by \cite{nature}.  This tail is less extended on the low-Li end when accounting for mass loss (i.e., the more realistic of the four models), which indicates that it reflects the initial conditions adopted in our models, combined with the fact that the less massive progenitors deplete Li more efficiently than the more massive ones.  

A large fraction of evolved GALAH stars do not have reported Li measurements, and it is natural to associate them with this predicted population. \cite{nature} contest this, on the grounds that the Li equivalent widths for the non-detections are in the detectable regime, and therefore the non-detections are not upper limits; instead they claim that they are not reported for other reasons. 

We find this explanation unconvincing for several reasons.  First, the Cannon methodology, used in the initial GALAH data release, is not designed to measure upper limits. However, it is clear from an examination of the observed pattern that a large number of the measurements must be Li non-detections.  Moreover, the Li detection limits, and measurements, are strong functions of effective temperature, with reported values all the way down to the detection limit at all values of T$_{\rm eff}$.  It is very unlikely that the true abundance distribution cuts off at exactly the same place where the lines become weak. The Li equivalent width can include other nearby weak lines, and a non-zero value does not require a detectable Li abundance.  Finally, open cluster samples of evolved giants have numerous true upper limits (see \citealt{magrini2021} for a most recent analysis based on data from the Gaia-ESO survey), and such stars must also exist in a field sample.  

We confirm these arguments with new GALAH data.  The latest GALAH Data Release (DR3,  \citealt{galahDR3}), uses the SME pipeline to infer Li abundances, and it does explicitly include upper limits.  We cross-matched the GALAH DR3 catalog with the \citet{zinn2021} K2 asteroseismic data.  We used the combination of the published GALAH effective temperatures and the seismic parameters from \citet{zinn2021} to infer masses, and we adopted the evolutionary states from that paper as well. 

In Figure \ref{fig:K2_galahdr3}, we show the results for RC and RGB stars.  More than 80\% of the measurements are upper limits in both the RC and RGB (top and bottom panels respectively). The Li abundances in the two groups are comparable, with the exception of a larger population of truly Li-rich stars in a minority of clump giants.  Most of the claimed detections are close to the detection limit and are marginal.  We caution that the distributions of masses is not the same as those in {\it Kepler}, so our simulation is not directly comparable to these data.  However, the large numbers of upper limits, along with the mean Li level in the clump, fit comfortably in the domain predicted by our simulations.

\section{Conclusions} \label{sect:conclusions}

A key lesson from our simulation is that Li depletion is a strong function of mass on the RGB, so a proper understanding of Li in evolved low-mass stars requires an understanding of the distribution of masses in the sample. Furthermore, a large number of very low Li abundances are expected for evolved stars, and a proper treatment of non-detections is therefore crucial.

We have modeled the expected distribution of Li abundances in the MS for the progenitors of present day field clump giants. Crucially, such progenitors have sufficiently different properties than today’s field red giants in what respects to the evolution of Li.  Our procedure then uses standard stellar models to simulate the distribution of Li abundances and the luminosities of such progenitors when arriving at the horizontal branch, in order to be compared with the GALAH Li data of field stars at that evolutionary state.  A full analysis should also attempt to match the metallicity distribution of both the RC and RGB, and to distinguish between high$-$ and low$-\alpha$ populations, but this is beyond the scope of this paper.  We adopted a stellar population model from the {\it Kepler} field, and different stellar populations in the GALAH footprint would impact our results. However, the central conclusions should be robust: the relative contributions of stars to the clump and giant branch depend sensitively on mass and composition, and we expect clump stars with detected Li to preferentially be produced in younger stellar populations relative to primarily old first ascent giants.

If there are no unknown mechanisms of Li production operating between the MS turnoff and the subsequent core helium burning phase, our simulated Li abundances should be comparable to those for low-mass field clump giants.  Our predicted distribution of Li abundances indeed matches the Li abundances of the bulk population of field RC stars as measured by GALAH.  Therefore, there is no evidence among clump giants of an unknown Li production mechanism occurring between the upper RGB and the horizontal branch for low-mass stars.

Furthermore, our model makes testable predictions: that the Li detections in the RC should be preferentially higher mass stars; that there are a large number of upper limits in the RC; and that the upper limits should be preferentially low mass stars. The data from \citet{carlberg2016} is very encouraging in this regard; they report Li detections for RC stars with masses from 1.6 to 2\msun\, in the range expected for our models.  A confirmed detection of Li in a large number of low mass RC stars (i.e., beyond the classical Li-rich ones) would indeed be evidence for Li production in at least some cases.  The limited overlap between our {\it K2} asteroseismic sample and GALAH DR3 consisted largely of upper limits or detections at a low level, along with a small fraction of true Li-rich stars. This is consistent with our model; a larger sample with more high-mass stars would be needed to see whether modest detection (A(Li) $\sim 1$) are associated with higher mass stars.

Finally, we stress that a crucial difference between the predictions of our simulations and the observed Li abundances of field RC stars is the presence in the data of stars with very high Li levels (up to A(Li) $\sim 3.5$ dex) that standard models cannot reproduce.  These are the well known Li-rich giants, which certainly remain an open problem, and are not within the scope of this paper.  However, the issue of what is the correct threshold for defining a giant star as normal or Li-rich needs to change and be incorporated in the mainstream urgently.  Specifically, classification schemes for Li-rich giants should not be based on a fixed threshold, but on a variable one that must depend on stellar mass \citep{claudia2016,trumpler2016}.  Otherwise, we will keep misinforming the efforts to study that truly challenging and longstanding problem in stellar evolution.

\section*{acknowledgments}

We are grateful to our referee, Sarah Martell, for an insightful and useful report that clarified our messages and improved this paper throughout.  We thank Con Deliyannis for providing us his data on lithium as a function of stellar mass for stars in NGC 6819, which we use to set the initial conditions of our simulations.  J.C. acknowledges support from the Agencia Nacional de Investigaci\'on y Desarrollo (ANID) via Proyecto Fondecyt Regular 1191366; and from ANID BASAL
projects CATA-Puente ACE210002 and CATA2-FB210003.  C.A.G acknowledges financial support from ANID FONDECYT Postdoctoral Fellowship 2018 Project 3180668.  JCZ is supported by an NSF Astronomy and Astrophysics Postdoctoral Fellowship under award AST-2001869.

\bibliographystyle{aasjournal}
\bibliography{references}

\begin{thebibliography}{}
\expandafter\ifx\csname natexlab\endcsname\relax\def\natexlab#1{#1}\fi

\bibitem[Aguilera-G{\'o}mez et al.(2020)]{li6} Aguilera-G{\'o}mez, C., 
Chanam{\'e}, J., \& Pinsonneault, M.~H.\ 2020, \apjl, 897, L20

\bibitem[{{Aguilera-G{\'o}mez}
  {et~al.}(2016{\natexlab{a}}){Aguilera-G{\'o}mez}, {Chanam{\'e}},
  {Pinsonneault}, \& {Carlberg}}]{trumpler2016}
{Aguilera-G{\'o}mez}, C., {Chanam{\'e}}, J., {Pinsonneault}, M.~H., \&
  {Carlberg}, J.~K. 2016{\natexlab{a}}, \apjl, 833, L24

\bibitem[{{Aguilera-G{\'o}mez}
  {et~al.}(2016{\natexlab{b}}){Aguilera-G{\'o}mez}, {Chanam{\'e}},
  {Pinsonneault}, \& {Carlberg}}]{claudia2016}
---. 2016{\natexlab{b}}, \apj, 829, 127

\bibitem[{{Aguilera-G{\'o}mez} {et~al.}(2018){Aguilera-G{\'o}mez},
  {Ram{\'\i}rez}, \& {Chanam{\'e}}}]{Lidesert}
{Aguilera-G{\'o}mez}, C., {Ram{\'\i}rez}, I., \& {Chanam{\'e}}, J. 2018, \aap,
  614, A55

\bibitem[Andrae et al.(2018)]{apsis} Andrae, R., Fouesneau, M., Creevey, O., et al.\ 2018, \aap, 616, A8

\bibitem[{{Balachandran}(1995)}]{balachandran1995}
{Balachandran}, S. 1995, \apj, 446, 203

\bibitem[{{Boesgaard} \& {Tripicco}(1986)}]{Lidip}
{Boesgaard}, A.~M., \& {Tripicco}, M.~J. 1986, \apjl, 302, L49

\bibitem[{{Buder} {et~al.}(2018){Buder}, {Asplund}, {Duong}, {Kos}, {Lind},
  {Ness}, {Sharma}, {Bland -Hawthorn}, {Casey}, {de Silva}, {D'Orazi},
  {Freeman}, {Lewis}, {Lin}, {Martell}, {Schlesinger}, {Simpson}, {Zucker},
  {Zwitter}, {Amarsi}, {Anguiano}, {Carollo}, {Casagrande}, {{\v{C}}otar},
  {Cottrell}, {da Costa}, {Gao}, {Hayden}, {Horner}, {Ireland}, {Kafle},
  {Munari}, {Nataf}, {Nordlander}, {Stello}, {Ting}, {Traven}, {Watson},
  {Wittenmyer}, {Wyse}, {Yong}, {Zinn}, {{\v{Z}}erjal}, \& {Galah
  Collaboration}}]{galah}
{Buder}, S., {Asplund}, M., {Duong}, L., {et~al.} 2018, \mnras, 478, 4513

\bibitem[{{Buder} {et~al.}(2021){Buder}, {Sharma}, {Kos}, {Amarsi},
  {Nordlander}, {Lind}, {Martell}, {Asplund}, {Bland-Hawthorn}, {Casey}, {de
  Silva}, {D'Orazi}, {Freeman}, {Hayden}, {Lewis}, {Lin}, {Schlesinger},
  {Simpson}, {Stello}, {Zucker}, {Zwitter}, {Beeson}, {Buck}, {Casagrande},
  {Clark}, {{\v{C}}otar}, {da Costa}, {de Grijs}, {Feuillet}, {Horner},
  {Kafle}, {Khanna}, {Kobayashi}, {Liu}, {Montet}, {Nandakumar}, {Nataf},
  {Ness}, {Spina}, {Tepper-Garc{\'\i}a}, {Ting}, {Traven},
  {Vogrin{\v{c}}i{\v{c}}}, {Wittenmyer}, {Wyse}, {{\v{Z}}erjal},
  {{\v{Z}}erjal}, \& {Galah Collaboration}}]{galahDR3}
{Buder}, S., {Sharma}, S., {Kos}, J., {et~al.} 2021, \mnras, 506, 150

\bibitem[{{Cameron} \& {Fowler}(1971)}]{cameron71}
{Cameron}, A.~G.~W., \& {Fowler}, W.~A. 1971, \apj, 164, 111

\bibitem[{{Carlberg} {et~al.}(2016){Carlberg}, {Cunha}, \&
  {Smith}}]{carlberg2016}
{Carlberg}, J.~K., {Cunha}, K., \& {Smith}, V.~V. 2016, \apj, 827, 129

\bibitem[Casey et al.(2019)]{casey2019} Casey, A.~R., Ho, A.~Y.~Q., Ness, M., et al.\ 2019, \apj, 880, 125

\bibitem[{{Cui} {et~al.}(2012){Cui}, {Zhao}, {Chu}, {Li}, {Li}, {Zhang}, {Su},
  {Yao}, {Wang}, {Xing}, {Li}, {Zhu}, {Wang}, {Gu}, {Luo}, {Xu}, {Zhang},
  {Liu}, {Zhang}, {Yang}, {Cao}, {Chen}, {Chen}, {Chen}, {Chen}, {Chu}, {Feng},
  {Gong}, {Hou}, {Hu}, {Hu}, {Hu}, {Jia}, {Jiang}, {Jiang}, {Jiang}, {Jin},
  {Li}, {Li}, {Li}, {Liu}, {Liu}, {Lu}, {Mao}, {Men}, {Qi}, {Qi}, {Shi},
  {Tang}, {Tao}, {Wang}, {Wang}, {Wang}, {Wang}, {Wang}, {Wang}, {Wang},
  {Wang}, {Wang}, {Wang}, {Wang}, {Wang}, {Xu}, {Xu}, {Yang}, {Yu}, {Yuan},
  {Yuan}, {Zhai}, {Zhang}, {Zhang}, {Zhang}, {Zhao}, {Zhou}, {Zhou}, {Zhu}, \&
  {Zou}}]{LAMOST}
{Cui}, X.-Q., {Zhao}, Y.-H., {Chu}, Y.-Q., {et~al.} 2012, Research in Astronomy
  and Astrophysics, 12, 1197

\bibitem[{{Deliyannis} {et~al.}(2019){Deliyannis}, {Anthony-Twarog},
  {Lee-Brown}, \& {Twarog}}]{ngc6819}
{Deliyannis}, C.~P., {Anthony-Twarog}, B.~J., {Lee-Brown}, D.~B., \& {Twarog},
  B.~A. 2019, \aj, 158, 163

\bibitem[{{Demarque} {et~al.}(2008){Demarque}, {Guenther}, {Li}, {Mazumdar}, \&
  {Straka}}]{demarque2008}
{Demarque}, P., {Guenther}, D.~B., {Li}, L.~H., {Mazumdar}, A., \& {Straka},
  C.~W. 2008, \apss, 316, 31

\bibitem[{{Gaia Collaboration} {et~al.}(2018){Gaia Collaboration}, {Brown},
  {Vallenari}, {Prusti}, {de Bruijne}, {Babusiaux}, {Bailer-Jones}, {Biermann},
  {Evans}, \& {Eyer}}]{dr2}
{Gaia Collaboration}, {Brown}, A.~G.~A., {Vallenari}, A., {et~al.} 2018, \aap,
  616, A1

\bibitem[Gao et al.(2019)]{gao2019} Gao, Q., Shi, J.-R., Yan, H.-L., et al.\ 2019, \apjs, 245, 33

\bibitem[Gao et al.(2021)]{gao2021} Gao, Q., Shi, J.-R., Yan, H.-L., et al.\ 2021, \apj, 914, 116

\bibitem[Gao et al.(2020)]{gao2020_Lidip} Gao, X., Lind, K., Amarsi, A.~M., et al.\ 2020, \mnras, 497, L30

\bibitem[{{Gilroy}(1989)}]{gilroy1989}
{Gilroy}, K.~K. 1989, \apj, 347, 835

\bibitem[{{Iben}(1965)}]{iben1965}
{Iben}, Icko, J. 1965, \apj, 141, 993

\bibitem[{{Izzo} {et~al.}(2015){Izzo}, {Della Valle}, {Mason}, {Matteucci},
  {Romano}, {Pasquini}, {Vanzi}, {Jordan}, {Fernand ez}, {Bluhm}, {Brahm},
  {Espinoza}, \& {Williams}}]{izzo2015}
{Izzo}, L., {Della Valle}, M., {Mason}, E., {et~al.} 2015, \apjl, 808, L14

\bibitem[{{Kumar} {et~al.}(2020){Kumar}, {Reddy}, {Campbell}, {Maben}, {Zhao},
  \& {Ting}}]{nature}
{Kumar}, Y.~B., {Reddy}, B.~E., {Campbell}, S.~W., {et~al.} 2020, Nature
  Astronomy, 4, 1059

\bibitem[{{Magrini} {et~al.}(2021){Magrini}, {Smiljanic}, {Franciosini},
  {Pasquini}, {Randich}, {Casali}, {Viscasillas Vazquez}, {Bragaglia}, {Spina},
  {Biazzo}, {Tautvaivsiene}, {Masseron}, {Van der Swaelmen}, {Pancino},
  {Jimenez-Esteban}, {Guiglion}, {Martell}, {Bensby}, {D'Orazi}, {Baratella},
  {Korn}, {Jofre}, {Gilmore}, {Worley}, {Hourihane}, {Gonneau}, {Sacco}, \&
  {Morbidelli}}]{magrini2021}
{Magrini}, L., {Smiljanic}, R., {Franciosini}, E., {et~al.} 2021, arXiv
  e-prints, arXiv:2108.11677

\bibitem[Martell et al.(2021)]{martell2021} Martell, S.~L., Simpson, J.~D., Balasubramaniam, A.~G., et al.\ 2021, \mnras, 505, 5340

\bibitem[Miglio et al.(2013)]{miglio2013} Miglio, A., Chiappini, C., Morel, T., et al.\ 2013, \mnras, 429, 423

\bibitem[{{Mori} {et~al.}(2021){Mori}, {Kusakabe}, {Balantekin}, {Kajino}, \&
  {Famiano}}]{neutrinos}
{Mori}, K., {Kusakabe}, M., {Balantekin}, A.~B., {Kajino}, T., \& {Famiano},
  M.~A. 2021, \mnras, 503, 2746

\bibitem[{{Olive} \& {Schramm}(1992)}]{spallation1992}
{Olive}, K.~A., \& {Schramm}, D.~N. 1992, \nat, 360, 439

\bibitem[{{Pace} {et~al.}(2012){Pace}, {Castro}, {Mel{\'e}ndez}, {Th{\'e}ado},
  \& {do Nascimento}}]{m67}
{Pace}, G., {Castro}, M., {Mel{\'e}ndez}, J., {Th{\'e}ado}, S., \& {do
  Nascimento}, J.~D., J. 2012, \aap, 541, A150

\bibitem[{{Pinsonneault}(1997)}]{pinsono1997}
{Pinsonneault}, M. 1997, \araa, 35, 557

\bibitem[{{Pinsonneault} {et~al.}(1989){Pinsonneault}, {Kawaler}, {Sofia}, \&
  {Demarque}}]{pinsono1989}
{Pinsonneault}, M.~H., {Kawaler}, S.~D., {Sofia}, S., \& {Demarque}, P. 1989,
  \apj, 338, 424

\bibitem[{{Pinsonneault} {et~al.}(2018){Pinsonneault}, {Elsworth}, {Tayar},
  {Serenelli}, {Stello}, {Zinn}, {Mathur}, {Garc{\'\i}a}, {Johnson}, {Hekker},
  {Huber}, {Kallinger}, {M{\'e}sz{\'a}ros}, {Mosser}, {Stassun}, {Girardi},
  {Rodrigues}, {Silva Aguirre}, {An}, {Basu}, {Chaplin}, {Corsaro}, {Cunha},
  {Garc{\'\i}a-Hern{\'a}ndez}, {Holtzman}, {J{\"o}nsson}, {Shetrone}, {Smith},
  {Sobeck}, {Stringfellow}, {Zamora}, {Beers}, {Fern{\'a}ndez-Trincado},
  {Frinchaboy}, {Hearty}, \& {Nitschelm}}]{apokasc}
{Pinsonneault}, M.~H., {Elsworth}, Y.~P., {Tayar}, J., {et~al.} 2018, \apjs,
  239, 32

\bibitem[{{Reeves} {et~al.}(1970){Reeves}, {Fowler}, \&
  {Hoyle}}]{spallation1970}
{Reeves}, H., {Fowler}, W.~A., \& {Hoyle}, F. 1970, \nat, 226, 727

\bibitem[{{Ryan} {et~al.}(2001){Ryan}, {Beers}, {Kajino}, \&
  {Rosolankova}}]{ryan2001}
{Ryan}, S.~G., {Beers}, T.~C., {Kajino}, T., \& {Rosolankova}, K. 2001, \apj,
  547, 231

\bibitem[{{Sackmann} \& {Boothroyd}(1992)}]{sackmann92}
{Sackmann}, I.~J., \& {Boothroyd}, A.~I. 1992, \apjl, 392, L71

\bibitem[{{Schwab}(2020)}]{flash}
{Schwab}, J. 2020, \apjl, 901, L18

\bibitem[{{Sestito} \& {Randich}(2005)}]{sestito}
{Sestito}, P., \& {Randich}, S. 2005, \aap, 442, 615

\bibitem[Sharma et al.(2016)]{sharma2016} Sharma, S., Stello, D., Bland-Hawthorn, J., et al.\ 2016, \apj, 822, 15

\bibitem[{{Shetrone} {et~al.}(2019){Shetrone}, {Tayar}, {Johnson}, {Somers},
  {Pinsonneault}, {Holtzman}, {Hasselquist}, {Masseron}, {M{\'e}sz{\'a}ros},
  {J{\"o}nsson}, {Hawkins}, {Sobeck}, {Zamora}, \&
  {Garc{\'\i}a-Hern{\'a}ndez}}]{shetrone2019}
{Shetrone}, M., {Tayar}, J., {Johnson}, J.~A., {et~al.} 2019, \apj, 872, 137

\bibitem[Smiljanic et al.(2018)]{smiljanic2018} Smiljanic, R., Franciosini, E., Bragaglia, A., et al.\ 2018, \aap, 617, A4

\bibitem[{{Somers} \& {Pinsonneault}(2016)}]{somers2016}
{Somers}, G., \& {Pinsonneault}, M.~H. 2016, \apj, 829, 32

\bibitem[{{Tajitsu} {et~al.}(2015){Tajitsu}, {Sadakane}, {Naito}, {Arai}, \&
  {Aoki}}]{tajitsu2015}
{Tajitsu}, A., {Sadakane}, K., {Naito}, H., {Arai}, A., \& {Aoki}, W. 2015,
  \nat, 518, 381

\bibitem[{{Vigroux} \& {Arnould}(1979)}]{novae1979}
{Vigroux}, L., \& {Arnould}, M. 1979, in Liege International Astrophysical
  Colloquia, Vol.~22, Liege International Astrophysical Colloquia, ed.
  A.~{Boury}, N.~{Grevesse}, \& L.~{Remy-Battiau}, 47--52

\bibitem[{{Zhang} {et~al.}(2021){Zhang}, {Shi}, {Yan}, {Li}, {Gao}, {Li},
  {Zhang}, {Liu}, {Bi}, {Zhao}, \& {Li}}]{zhang2021}
{Zhang}, J., {Shi}, J.-R., {Yan}, H.-L., {et~al.} 2021, arXiv e-prints,
  arXiv:2108.13030

\bibitem[{{Zinn} {et~al.}(2021){Zinn}, {Stello}, {Elsworth}, {Garc{\'\i}a},
  {Kallinger}, {Mathur}, {Mosser}, {Hon}, {Bugnet}, {Jones}, {Reyes}, {Sharma},
  {Sch{\"o}nrich}, {Warfield}, {Luger}, {Vanderburg}, {Kobayashi},
  {Pinsonneault}, {Johnson}, {Huber}, {Buder}, {Joyce}, {Bland-Hawthorn},
  {Casagrande}, {Lewis}, {Miglio}, {Nordlander}, {Davies}, {De Silva},
  {Chaplin}, \& {Silva Aguirre}}]{zinn2021}
{Zinn}, J.~C., {Stello}, D., {Elsworth}, Y., {et~al.} 2021, arXiv e-prints,
  arXiv:2108.05455

\end{thebibliography}

\end{document}